# Stability of Supported Pd-based Ethanol Oxidation Reaction Electrocatalysts in Alkaline Media

*Tuani C. Gentil[a,b], Maria Minichova[a,c], Valentín Briega-Martos[a], Victor S. Pinheiro[b], Felipe M. Souza[b,d], João Paulo C. Moura[b], Júlio César M. Silva[e], Bruno L. Batista[b], Mauro C. Santos[b*], Serhiy Cherevko[a*]*

[a] *Helmholtz-Institute Erlangen-Nürnberg for Renewable Energy (IET-2), Forschungszentrum Jülich, Cauerstr. 1, 91058 Erlangen, Germany*

[b] *Centro de Ciências Naturais e Humanas (CCNH), Universidade Federal do ABC (UFABC). Rua Santa Adélia 166, Bairro Bangu, 09210-170, Santo André - SP, Brazil.*

[c] *Department of Chemical and Biological Engineering, Friedrich-Alexander-Universität Erlangen-Nürnberg, Cauerstr. 1, 91058 Erlangen, Germany*

[d] *Instituto Federal de Educação, Ciência e Tecnologia Goiano, BR-153, Km 633, Zona Rural, 75650-000, Morrinhos, Goiás, Brazil*

[e] *Universidade Federal Fluminense (UFF), Instituto de Química, Grupo de Eletroquímica e Materiais Nanoestruturados, Campus Valonguinho, 24020-141, Niterói, Rio de Janeiro, Brazil*

*Corresponding Authors:*

*E-mail: pdrmcsa@gmail.com*
*E-mail: s.cherevko@fz-juelich.de*

**ABSTRACT**

This study evaluates the dissolution of the supported electrocatalysts Pd/C, PdSn/C, PdNb/C, and PdFe$_3$O$_4$/C during ethanol oxidation reaction for ADLFC applications. A scanning flow cell (SFC) coupled to an inductively coupled mass spectrometry (online ICP-MS) is used to assess the dissolution stability in a broad potential window. Accelerated stress tests with and without ethanol are developed using a rotating disk electrode (RDE) with dissolution products analysis by ex-situ ICP-MS. Potential profiles simulating those experienced by the catalyst during regular fuel cell operation were used. Sn and Fe catalysts demonstrate improved activity and stability compared with the material with Pd alone. For these reasons, PdSn/C and PdFe$_3$O$_4$/C are suitable for ADLFC applications. Severe Nb dissolution destabilizes Pd, increasing its leaching. This work demonstrates that while additional metals and oxides can improve the alcohol oxidation kinetics of Pd, these additives' dissolution stability must already be considered at the catalyst design stage.

# 1. INTRODUCTION

Due to population growth, new energy sources are constantly being explored to meet the world's increasing demands. Fuel cells are noteworthy, as they comprise devices capable of converting chemical energy into electrical energy, presenting high efficiency and low pollutant emissions [1–3]. In this context, $H_2$-$O_2$ fuel cells, despite having a thermodynamic efficiency of over 80% [1,2], still face challenges due to the explosive nature of hydrogen, which makes the transport and storage of the fuel difficult [4,5]. Another disadvantage of hydrogen is its low volumetric energy density (5.6 MJ $dm^{-3}$ compressed to 700 bar), lower than gasoline, methanol, and ethanol [6]. Viable alternatives in this regard comprise direct liquid fuel cells (DLFCs), which can operate in both acid or alkaline media and allow the use of fuels such as ethanol, glycerol, ethylene glycol, and more recently, acetol (1-hydroxy-2-propanone) [7–13].

Palladium-based electrocatalysts supported on carbon have shown promising results in terms of current density and power in Alkaline Direct Liquid Fuel Cells (ADLFCs), also presenting lower onset potential compared to Pt electrocatalysts and higher peak current densities for the ethanol oxidation reaction (EOR) in alkaline media [2,14]. Although palladium is an outstanding catalyst in alkaline media, the slow kinetics of the anodic alcohol oxidation reaction and the poisoning of the electrode surface by active intermediates such as $CH_3CH_2O_{ad}$ and $CH_3CO_{ad}$, generated from incomplete ethanol oxidation, remain significant obstacles in the development of electrocatalysts for Alkaline Direct Ethanol Fuel Cells (ADEFC) [3,15,16].

Besides carbon-supported, Pd co-catalysts based on transition metals and their oxides have been studied in recent decades to explore their synergistic effects towards the electrooxidation of organic molecules and reduce the cost of metallic loads resulting from Pd [17–22]. Co-catalysts can also facilitate the oxidation of intermediate compounds, providing oxygenated species, favoring bifunctional mechanisms, and reducing catalytic poisoning [23–25]. In addition, auxiliary metals can also promote the electronic effect that occurs when other metals modify the electronic structure of Pd with different d-orbital fillings [26–28]. The shift in Pd binding energy and d-band center interferes with the interaction between Pd and the small organic molecules [29]. Adding Sn, Fe, and Nb to Pd has been reported to be especially promising in improving the electrocatalytic activity of the resultant composite catalysts. Hence, we provide a brief literature overview below on using these three metal oxides with Pd towards EOR.

Pinheiro *et al.* [7] studied PdSn-based materials for EOR in alkaline media at different mass ratios. They verified that oxygenated species, vacancies, and defects detected in the structures of $Pd_xSn_y$/Vulcan XC-72 catalysts resulted in materials with higher electrocatalytic activities when compared to simple Pd/C, in addition to the presence of oxophilic Sn species, which improve their ability to remove adsorbed CO.

Studies show that the presence of Fe can also favor EOR when used in conjunction with noble metals such as Pd and Pt [30–33]. Wang *et al.* [32] synthesized FePd-nanoalloys deposited on gamma $\gamma$-$Fe_2O_3$, FePd-$Fe_2O_3$, anchored on carboxyl multi-walled carbon nanotubes (MWNTs) and showed an approximately 3.65 fold increase in current density peak (normalized by mass) for EOR compared to Pd/MWNTs. In addition to the alloy approach, some authors use the presence of iron oxides in conjunction with Pd, which may reduce the energy barriers for ethanol oxidation due to an improvement in the electronic properties of the electrocatalyst [31].

Niobium has been recently applied as a co-catalyst for ethanol and glycerol oxidation reactions, and the findings indicated that $Pd_xNb_y$/C anode electrocatalysts applied to alkaline direct glycerol fuel cells resulted in a maximum power density of 27 mW $cm^{-2}$ using PdNb/C, versus 18.11 mW $cm^{-2}$ in alkaline DEFC using the same mass ratio material [10,23].

The electrochemical and operational results of the literature show that bimetallic electrocatalysts are promising. However, the performance of an ADLFC can drop considerably according to long-term durability tests, making it extremely important to understand the degradation processes that may occur during operation [34]. The dissolution of metals is directly related to the durability of these devices and has been extensively studied in recent years in acid and alkaline media. Thus, the increasing application of Pd-based materials alongside co-catalysts in ADLFC and the promising catalytic activity results in the need for dissolution studies of Pd and co-catalysts in alkaline media in conditions close to actual operation [12,35–37].

Online and long-term stability studies will allow for a better understanding of the life cycle of a fuel cell employing this type of electrocatalyst since its degradation can significantly impact the fuel cell performance and costs [38–40]. In this context, investigations are being conducted by applying accelerated stress testing (AST) to determine material durability. AST can be performed in situ, using a complete fuel cell, or ex-situ, with a half-electrochemical cell under controlled potential, facilitating isolated studies without external factor interferences during cell operation [41].

The innovative scanning flow cell technique and an inductively coupled plasma mass spectrometer (online SFC-ICP-MS) allow real-time metal dissolution tracking during the applied electrochemical protocol. The principle of this technique is a system composed of a three-electrode electrochemical half-cell where the reaction takes place, coupled to an ICP-MS capable of detecting multi-elements dissolved in the electrolyte at a wide linear quantification range. In this system, electrochemical parameters are varied, simulating the conditions applied to fuel cells to obtain information about how they affect the electrocatalyst stability [39,42]. Several previous online ICP-MS studies focused on the dissolution of Pd. Pizzutilo *et al.* [43] conducted studies regarding Pd dissolution in acidic media employing online ICP-MS, demonstrating that potential sweep rate, upper potential limit (UPL), and electrolyte composition directly influence Pd dissolution. The authors compared the dissolution of bulk polycrystalline Pd (poly-Pd) with Pd/C materials and verified that transient dissolution is promoted mainly by one anodic and two cathodic contributions. However, the growing demand for Pd application in alkaline media also requires studies of Pd-based electrocatalysts destined for ADLFC.

This study presents the effect of an alkaline electrolyte and organic biofuel (ethanol) on the stability of the Pd/C, PdSn/C, PdNb/C, and $PdFe_3O_4$/C electrodes. Fundamental electrocatalyst stability studies were first carried out using the online SFC-ICP-MS technique in the absence and presence of ethanol. Electrocatalyst stability was also investigated through ASTs using a rotating disk electrode (RDE) and offline ICP-MS measurements to mimic actual ADLFC operating conditions. Thus, the main aim of this study is to evaluate the dissolution stability of both Pd and supports in Pd-based electrocatalysts destined for ADLFC anodes, highlighting that the combination of dissolution studies by offline ICP-MS measurements close to actual ADLFC conditions, fundamental on-line dissolution studies by SFC, are not common in the literature, which is the main innovation of this work.

2. EXPERIMENTAL
*1.1. Preparation of the electrocatalysts*

Pd-nanoparticles and PdSn/C were synthesized by the chemical reduction method via sodium borohydride as described by Pinheiro et al. [6], PdNb/C was synthesized by the sol-gel method as described by Moura *et al.* [23], and the nanostructures were supported on Vulcan XC-72 carbon (Cabot Corporation) [7,8,23,24]. The synthesis of nano-octahedral Fe$_3$O$_4$ was performed through the hydrothermal route [44], with PdFe$_3$O$_4$/C formed by the chemical reduction method and hydrothermal route. The electrocatalysts were synthesized considering approximately 85 % by mass of Vulcan XC-72 and a 1:1 ratio of metals by mass of the Pd co-catalyst. The reference electrocatalyst was palladium, 20% on activated carbon powder, standard, Alfa Aesar Company (United States),

*1.2. Physical Characterisation*

Transmission electron microscopy (TEM) images of the electrocatalysts recorded using a JEOL JEM 2100 microscope operating at 200 kV determined Nanoparticle morphology and size. The electrocatalysts were physically characterized employing X-ray diffraction (XRD), where the electrocatalyst structures were studied by XRD patterns using a D8 Focus diffractometer (Bruker AXS) with CuKα radiation, operating in continuous scan mode (2° min$^{-1}$) from 10° to 90°. The electrocatalysts' element distribution and chemical composition were estimated through scanning electron microscopy (SEM) images and energy dispersive spectroscopy (EDS) spectra, respectively, employing a JEOL JSM SEM-6010LA equipment. The amount of Pd and Fe in the electrocatalysts was determined using an inductively coupled plasma mass spectrometer (ICP-MS, Agilent 7900, Hachioji, Japan) operated with high-purity argon (99.9999%, White Martins, Brazil).

*1.3. Electrochemical experiments*

Accelerated stress tests (AST) in alkaline media were performed in a PTFE electrochemical cell composed of a double compartment Ag/AgCl reference electrode (Ag/AgCl (3 M KCl, Metrohm), a glassy carbon rod as the counter electrode (HTW Sigradur G), and glassy carbon (GC) rotating disk electrode (RDE) (Modulated Speed Rotator (MSR), Pine Research, geometric surface area A$_{geo}$ = 0.196 cm$^2$) [12,45–48]. Before each measurement, the GC electrode was polished with MD-Mol paste (water-based diamond suspension; particle size of 3 μm, Struers) and washed with ultra-pure water. Electrocatalyst dispersions (approximately 5 mg of material per mL) were prepared in a mixture of ultrapure water: isopropanol v/v (80:20 respectively). Nafion perfluorinated 25% of the catalyst resin solution (Sigma Aldrich, five wt %, 28.6 μL) was added, and the pH was subsequently adjusted to 11 using KOH (Merck 1.0 M). The dispersion was sonicated for 20 minutes using a sonic horn at 40% intensity (Branson SFX 150). Sonication was activated for 4 s and interrupted for a 2 s interval utilizing an ice bath during sonication. After sonication, 19.6 μL aliquots of ink were drop-cast on the RDE and dried at room temperature at a rotation speed of 60 rpm. The measurements were carried out using 1.0 M KOH (Fe, Ni, Cu≤0.0005%, Merck Emsure) electrolyte prepared with ultrapure water (Merck Millipore Milli-Q system, 18.2 MΩ cm TOC < 2.0 p.p.b.) and bubbled with Ar (99.999%, Air Liquide Deutschland GmbH) in both the absence and presence of 1.0 M ethanol (Merck KGaA, absolute for analysis) for the EOR tests. The electrolytes were collected before and after 5000 cycles, using a potential window of 0.07 to 0.5 $V_{RHE}$ and applying a 500 mVs$^{-1}$ scan rate. The pH of the electrolyte solutions was measured and used to convert the measured potentials versus the Ag/AgCl reference electrode ($E_{applied}$) to the reversible hydrogen electrode (RHE) scale ($E_{RHE}$), employing the following equation [12,49], where $E_{Ag/AgCl}$ is the potential of the Ag/AgCl Reference electrode:

$$E_{RHE} = E_{applied} + E_{Ag/AgCl} + 0.0591 \times pH \quad\quad (1)$$

The value of the Ag/AgCl reference electrode was carefully checked before and after every measurement to ensure no fluctuations occurred during the experiments.

### 1.4. Preparation of thin film catalyst layers for SFC-ICP-MS measurements

Thin Pd-based material films were prepared on a glassy carbon (GC) substrate (SIGRADUR G, HTW 25 cm$^2$) by drop-casting [12]. The GC substrates were previously ground with SiC grinding paper applying 300 N force and 200 rpm, using four different grain sizes: 220, 1000, 2000, and 4000 (corresponding to 68, 18, 10, and 5 μm grain size, respectively). The GC electrode was polished using the same device on an MD-Mol (Cat. number: 40500079) polishing pad using DiaPro MD-Mol paste (water-based diamond suspension; particle size of 3 μm, Struers) at 150 N and 200 rpm, and the sample holder was also counter-rotated at 150 rpm for 5 min. The flat GCs were cleaned with Kimwipes and isopropanol, rinsed with Milli-Q water, and dried. 0.3 μL aliquots of the inks prepared as described in Section 2.4 were pipetted and drop-cast on the GC substrate, reaching approximately 20 μg cm$^{-2}$ total metal loading. The spots were then air-dried at room temperature. The diameter and quality of each spot were determined using a laser scanning microscope (Keyence VK-X250), ranging between 600 and 800 μm. All electrochemical and dissolution data were normalized against the particular geometric surface area.

*1.5. Stability measurements using on-line ICP-MS*

Stability measurements were performed employing a microelectrochemical scanning flow cell (SFC) coupled to an inductively coupled plasma mass spectrometer (ICP-MS/ Nexion 300X, PerkinElmer) [12,39]. A GC rod (HTW Sigradur G) and a commercial double compartment Ag/AgCl (3 M KCl, Metrohm) attached to the SFC inlet and outlet tubes, respectively, were used as counter and reference electrodes. The drop-casted spots on the glassy carbon GC substrate were used as working electrodes. The potentiostat (Gamry Reference 600), electrolyte, gas flow, and SFC components are automatically controlled using in-house LabVIEW software. A 0.05 M KOH solution (Fe, Ni, Cu≤0.0005%, Merck Emsure) was used as the electrolyte in all experiments, employing 0.05 M ethanol (Merck KGaA, absolute for analysis) as fuel and ultrapure water (Merck Millipore Milli-Q system, 18.2 MΩ cm TOC < 2.0 p.p.b.). The concentrations of the supporting electrolyte and ethanol were limited to 0.05 M due to technical limitations of the ICP-MS. The experiments were carried out with the electrolyte in the presence and absence of fuels, and the freshly prepared electrolyte was pumped through the microelectrochemical cell and over the working electrode, which was directly introduced into the ICP-MS system. The equipment consists of a cyclonic spray chamber and a nebulizer. The power for the plasma is held at 1300 W with a gas flow of 15 L min$^{-1}$. The electrolyte was mixed with an internal standard using a Y-connector downstream of the SFC and introduced online into the ICP-MS [39] with a ca. 200 μL·min-1 flow rate with continuous argon purging. Calibration was performed from Pd, Sn, Nb, and Fe solutions (Certipur®, Merck), while ten μg·L-1 Rh (for Pd and Sn), Y (for Nb), and Ge (for Fe) were used as internal standards. Total quantities of dissolution were obtained via integration of the transient dissolution profiles.

3. **RESULTS AND DISCUSSION**
*3.1. Physical electrocatalyst characterization*

The commercial and synthesized Pd/C [7,8], PdSn/C [7], and PdNb/C [23] used in this work were already investigated in our previous studies [7,8,23]. Hence, the corresponding physical characterization results are discussed briefly below. On the other hand, a more detailed analysis is performed for the newly synthesized PdFe$_3$O$_4$/C. A set of techniques, including XRD, ICP-MS, and HR-TEM, was applied to obtain information on the materials' morphology, composition, and structure. The XRD patterns of all electrocatalysts are displayed in the Supporting Information (SI) (**Fig. S1**). The elemental compositions of the electrocatalysts obtained by ICP-MS, confirming the nominal mass ratios, are shown in **Tab. S1**.

Two Pd/C catalysts were studied in this work: (a) Commercial Pd/C from Alfa Aesar (denoted as Pd/C$_{AA}$ below) used as a reference and (b) synthesized Pd/C using a similar synthesis procedure as for other catalysts. HR-TEM images of both materials are shown in SI (**Fig. S2a** and **S2b**, respectively). The average particle size is 4.9 ± 1.3 nm and 8.7 ± 1.6 nm for Pd/C$_{AA}$ and Pd/C, respectively [7]. While the nanoparticles are uniformly distributed in Pd/C$_{AA}$ material, some agglomerations are present for the synthesized Pd/C electrocatalyst. Some differences can be discerned in the morphology of the particles since the (220), (311), and (222) reflections observed for the synthesized Pd/C material do not appear for the Pd/C$_{AA}$ reference.

Previous EDS and ICP-MS results indicate that the PdSn/C presents an elemental composition close to the initial nominal mass ratio, with averages of Pd$_{1.0}$ and Sn$_{1.1}$ [7]. The electrocatalyst XRD patterns of the Pd/C and PdSn/C demonstrate the presence of metallic Pd (PDF 5-681) with a cubic fcc(ccp)-Cu structure, as confirmed by the (111) and (200) reflections at 40° and 46°, respectively, when compared with the Pd/C$_{AA}$. The XRD analysis also suggests the formation of a Pd-Sn alloy in the case of the PdSn/C material, indicated by reflections (001), (100), and (011) at 30°, 32° and 45°, and the presence of SnO$_2$. The crystallite size for the PdSn/C, obtained using Scherrer's equation, was 5.1 ± 2.1 nm. The net parameters of the Pd (200) peak were calculated, which suggests net lattice expansion, indicating the formation of a Pd-Sn alloy. The HR-TEM images (**Fig. S2c**) demonstrate greater nanoparticle homogeneity on the carbon support for the PdSn/C compared to the pure Pd/C (**Fig. S2b**). Furthermore, the average size of the Pd nanoparticles in the PdSn/C was 9.1 ± 2.5 nm [7]. The observed discrepancy between the XRD and HR-TEM results can be explained by forming particles composed of several grains.

Regarding the binary PdNb/C, the nominal Pd and Nb mass concentrations analysis by ICP-MS indicated 8.8 ± 0.6% and 13.4 ± 3.6% Pd and Nb electrocatalyst content, respectively [23]. The XRD results shown in **Fig. S1** suggest that the evaluated catalysts are a mixture of metals and oxides of the corresponding elements. Thus, unlike PdSn/C, PdNb/C consists of separate Pd and Nb oxide phases without forming Pd-Nb alloys. The average crystallite size calculated by the Scherrer equation is 4.1 ± 1.1 nm.

HR-TEM images (**Fig. S2d**) indicate a random nanoparticle distribution on the Vulcan XC72 carbon support, with mean diameters of 6.4 ± 3.0 in the PdNb/C, but with some aggregations [23].

The shown in **Fig. S1** XRD patterns of the PdFe$_3$O$_4$/C indicate the presence of metallic Pd (PDF 5-681) with a cubic fcc(ccp)-Cu structure, as confirmed at 40° and 46°, by reflections (111) and (200), respectively. The presence of amorphous carbon was confirmed by a broad (002) reflection at about 25°. The plane of reflection (101) at 34° suggests the presence of PdO with a tetragonal crystal structure compared to the reference diffraction patterns (PDF 01-075-0584) [7]. Diffraction Fe$_3$O$_4$ peaks are observed at 18.0°, 30.2°, 35.5°, 43.3°, 57.2°, 63.0°, 74.0° referring to reflection planes of (111), (220), (311), (400), (511), (440) and (533), respectively. The PDF 19-629 standards were employed, and the diffraction peaks can be indexed to a cubic magnetite structure phase (fcc, space group Fd-3 m) to identify Fe$_3$O$_4$ phases and crystalline systems [31,32]. The interplanar distance was calculated using Bragg's law and the lattice Pd peak (200) parameter by applying Scherrer's equation in the PdFe$_3$O$_4$/C diffractogram, and a lattice parameter variation was observed. The average crystallite size calculated by Scherrer's equation was 5.1 ± 0.7 nm. **Fig. S3** displays the HR-TEM image of the PdFe$_3$O$_4$/C with Pd nanoparticles, Fe$_3$O$_4$ nano-octahedra, and carbon Vulcan XC72, confirming compositions and controlled morphologies. The average Pd nanoparticle size is 5.1 ± 0.9 nm, close to those obtained from XRD standards.

Differences in particle morphology, size, and distribution (aggregation) between the different electrocatalysts can affect their activity and dissolution behavior. The following sections will discuss these issues, together with the presented results.

### 3.2. Online SFC-ICP-MS dissolution

A broad potential window, mimicking the regular anode operation, fuel starvation, and cell-off events, was selected to assess the dissolution stability of Pd and alloying/mixing secondary metal elements. The main objective of the online SFC-ICP-MS studies was to measure the onsets and extents of Pd, Sn, Nb, and Fe dissolution, mapping the stability of the Pd/C, PdSn/C PdNb/C, and PdFe$_3$O$_4$/C in the absence and presence of ethanol. The dissolution profiles of the electrocatalysts in a 0.05 M KOH in the absence of ethanol are shown in **Fig. 1**. The observed noise in the measured dissolution profiles is in line with the results from Kormányos *et al.* [12], who conducted comparative studies in acidic and alkaline media and attributed higher noise to the presence of KOH.

To track how anodic and cathodic processes influence the dissolution, subsequent CVs at a scan rate of 5 mV s$^{-1}$ with continuously increasing upper potential limits (UPLs) of cycles going to 0.3 V$_{RHE}$, 0.5 V$_{RHE}$, 0.7 V$_{RHE}$, 0.9 V$_{RHE}$, 1.0 V$_{RHE}$, and 1.3 V$_{RHEs}$ were recorded. The CVs were preceded and followed by potential holds at 0.07 V$_{RHE}$. The first dissolution features are visible during the potentiostatic hold at 0.07 V$_{RHE}$ before CV recording, which emerged when the SFC contacted the working electrode (contact peak). This signal usually comes from the cathodic reduction of native surface oxides previously present in the sample [50]. The Pd dissolution contact peak was more pronounced in the Pd/C$_{AA}$ than in the synthesized materials. This result can be explained by the smaller particle size of Pd/C$_{AA}$, as smaller particles are more prone to oxidation in the air and the lower particle agglomeration. Comparing the dissolution of Pd and alloying/mixing elements during the initial catalyst contact with the electrolyte, the values for Sn and Nb were higher than that for Pd, while no Fe dissolution was observed. This observation indicates that in the support systems, the stability of the primary catalyst and an additional element must be studied simultaneously.

During CVs, a small Pd dissolution peak starts to evolve in the fourth CV with 0.7 V$_{RHE}$ UPL for Pd/C and PdNb/C and a higher UPL of 0.9 V$_{RHE}$ for the Pd/C$_{AA}$, PdSn/C and PdFe$_3$O$_4$/C. In this case, the lower onset of dissolution for the Pd/C compared to the Pd/C$_{AA}$ must arise from differences in the morphology and uniformity of the nanoparticles, as suggested by the XRD analysis [7]. The anodic and cathodic dissolution increases with each new cycle up to the UPL 1.3 V$_{RHE}$. While Sn dissolution was measured during the protocol, dissolution peaks for Nb are difficult to distinguish due to the intense contact peak and tailing. When comparing the different synthesized catalysts, the PdFe$_3$O$_4$ shows considerably less Pd dissolution at low potentials, suggesting a stabilizing effect by the presence of Fe. The onset potential of dissolution also seems to be shifted to more positive values in the case of PdSn/C. Still, in this case, the dissolution amounts for UPL = 0.9 V is not so different, and therefore, the stabilising effect is not so clear in this case.

The study was continued by testing catalyst stability in the presence of ethanol (0.05 M ethanol in 0.05 M KOH), and the results are shown in the support information (**Fig. S4** and **Fig. S5**). Qualitatively, the dissolution behavior of all catalysts is similar to that observed in the fuel-free electrolyte. The critical difference is that, except for PdSn/C, the small dissolution features can be seen at lower UPLs. This fact implies that higher dissolution might be observed in the presence of ethanol at lower anodic potentials. No significant differences can be discerned for the secondary elements in the presence of ethanol. The onset potential calculated from the peak with 1.0 V$_{RHE}$ UPL is essentially the same for all samples and is not affected by ethanol (**Tab. S2**) [12].

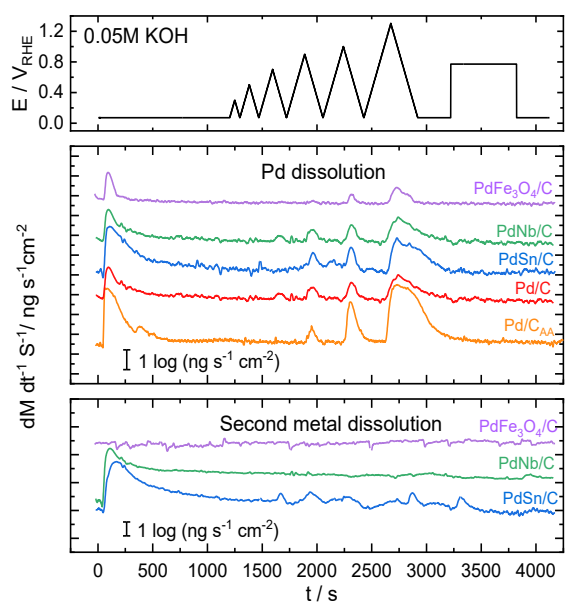

**Fig. 1**. Dissolution rates of Pd-based electrocatalysts recorded during the electrochemical protocol applying a 5 mV s$^{-1}$ scan rate in a 0.05 KOH solution. Logarithmised data was used for better visibility of dissolution peaks.

To quantify the amounts of dissolved Pd and secondary metals during CVs, the presented dissolution profiles in **Fig. 1** and **Fig. S4** were integrated with the results shown in **Fig. 2**. For clarity, only one CV with UPL of 1.0 V$_{RHE}$ was selected for the integration. Considering error bars, there is no significant difference in Pd dissolution in electrolytes with and without ethanol. Due to the limitations of ICP-MS, however, relatively low concentrations of ethanol and KOH were used in these experiments. Hence, while no influence of ethanol on dissolution was observed, we cannot exclude that such dependence exists at higher concentrations. Considering the limitations of on-line ICP-MS, i.e., short protocols, dissolution observation at relatively high anodic potential, mild model electrolytes, and prolonged RDE experiments were planned and executed with results discussed in the following sections.

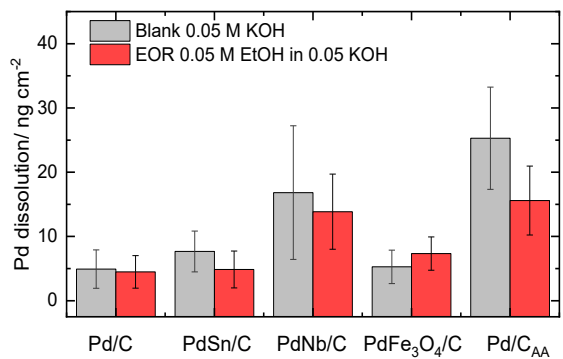

**Fig. 2.** Amounts of dissolved Pd, integrated curves of 5$^{th}$ CV (UPL 1 V$_{RHE}$) from Fig. 1 and Fig. S4.

*3.3. Electrochemical behavior of the catalysts before and after AST*

Considering an operating fuel cell, anode catalysts are unlikely to be exposed to potentials higher than 1.0 V$_{RHE}$ during regular operation, including cell-off periods with oxygen penetration into the anode compartment [12]. Even lower anodic potentials are expected during load cycles without fuel starvation. Thus, Pd dissolution at relatively high anodic potentials in the on-line ICP-MS study (**Fig. 3** and **4**) does not necessarily indicate low catalyst stability in real devices. On the other hand, the absence of catalyst dissolution at low potentials during short protocols does not guarantee any dissolution on longer time scales. Hence, ASTs were performed herein at a potential range from 0.07 to 0.5 V$_{RHE}$ and a scan rate of 500 mVs$^{-1}$ in the absence and presence of fuel. The UPL was chosen to be 0.5 V$_{RHE}$ since it is located at the beginning of the ethanol oxidation peak, as shown below. Higher potentials would result in impractical low cell voltages, considering oxygen reduction reaction (ORR) on the cathode.

**Fig. 3** displays the CVs before and after the 5000 CV cycles AST in 1.0 M KOH blank electrolyte. Here, a vast potential window from 0.07 to a relatively high anodic potential of 1.27 V$_{RHE}$ is chosen to study how AST affects both hydrogen and oxygen potential regions of Pd CVs. PdO formation occurs on the surface of the electrocatalyst during the direct scan, starting roughly at 0.7 V$_{RHE}$. At the same time, the reduction of the formed oxide initiates at ca. 0.75 V$_{RHE}$ during the reverse scan [7].

We first discuss the CVs of as-synthesized materials, as significant differences were found when comparing different materials. The most striking observation is that the current from Pd/C$_{AA}$ is several times higher than for all other catalysts, implying a higher surface area due to lower particle size and better homogeneity. This result explains the higher dissolution amount of Pd/C$_{AA}$ in the on-line ICP-MS study. All other catalysts, except PdNb/C, reveal similar currents in hydrogen and oxygen regions. The relatively low PdOx reduction peak of PdNb/C may imply surface poisoning, lower surface area because of aggregation, or potential dependent conductivity change between Pd and Nb oxide, as shown for the Pd/CeO$_X$ system [14]. Comparing the CVs before and after the AST, no significant differences were observed in the hydrogen region for all catalysts. On the other hand, the AST resulted in a slight decrease in the PdOx reduction peak for Pd/C$_{AA}$ and PdFe$_3$O$_4$/C, while other materials did not reveal any differences.

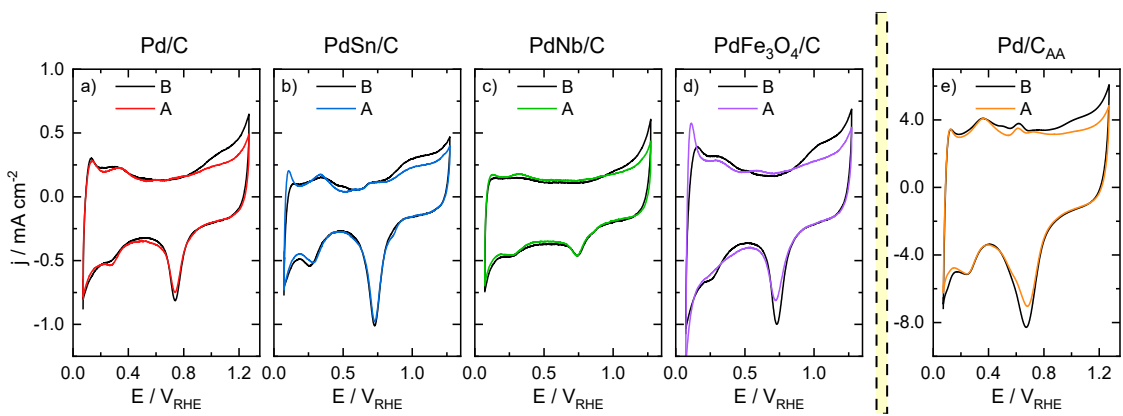

**Fig. 3**. Cyclic voltammograms recorded for Pd/C (a), PdSn/C (b), PdNb/C (c), PdFe$_3$O$_4$/C (d), and Pd/C$_{AA}$ (e) at a scan rate of 20 mVs$^{-1}$, in 1.0 M KOH before (B) and after (A) 5000 AST cycles. The experiments were performed with RDE configuration and constant Ar purge.

The same AST was performed in the electrolyte in the presence of 1 M ethanol. **Fig. 4** shows the anodic direction scans of the CVs before and after AST. Except for the Pd/C$_{AA}$, which displayed an 11% current decrease after the AST, the EOR activity of all studied catalysts improved. The ascending order is PdNb/C < Pd/C < PdFe$_3$O$_4$/C < PdSn/C. The PdSn/C resulted in 77 mA cm$^{-2}$ for EOR, followed by PdFe$_3$O$_4$/C at 38 mA cm$^{-2}$ after AST.

In the present work, the measured currents are normalized to the geometric area, and the electrocatalyst films were prepared using the same total mass of material. Since in the work by Souza *et al.* [10], the results are normalized by the mass of Pd, this would be the reason why, in that work, the current density results are similar between Pd/C and PdNb/C, while here, the current density of PdNb/C is ca. half of the value for Pd/C. The higher current density presented here for the Pd/C$_{AA}$ compared to the synthesized Pd/C must be due to the larger particle size and higher agglomeration of the latter. As indicated by stability studies, the lower current density observed for PdNb/C may also result from Pd dissolution in this electrocatalyst. In contrast, the higher activities for PdSn/C and PdFe$_3$O$_4$ align with the secondary metal's enhancing effect, in agreement with their good particle distribution.

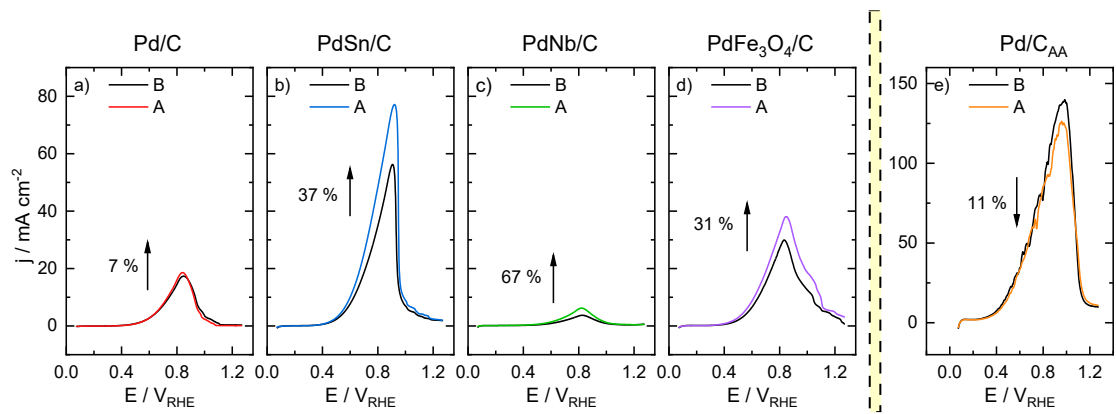

**Fig. 4.** Positive-going scan recorded for Pd/C (a), PdSn/C (b), PdNb/C (c), PdFe$_3$O$_4$/C (d), and Pd/C$_{AA}$ (e), applying a 20 mV s$^{-1}$ scan rate, in 1.0 M EtOH and 1.0 M KOH before (B) and after (A) 5000 AST cycles. The experiments were performed with RDE configuration and constant Ar purge.

### 3.4. Dissolution during the AST

The accumulated dissolved species in the electrolytes resulting from the ASTs were analyzed using ex-situ ICP-MS. To account for the contact peak dissolution, aliquots of the electrolytes were taken before and after the AST, while the fresh electrolyte was considered metal-free. The dissolution percentages of all the metals are detailed in **Tab. S3** (1.0 M KOH) and **Tab. S4** (1.0 M KOH + 1.0 M EtOH). **Fig. 5** depicts the relative percentage of Pd and secondary metal dissolution in 1.0 M KOH **(a)** and 1.0 M KOH + 1.0 M EtOH **(b)**.

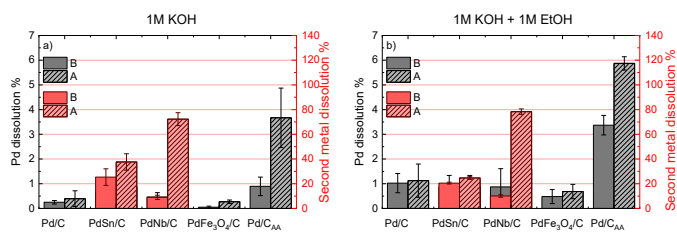

**Fig. 5**. Total dissolution from catalyst layer of Pd (left axis) and second metal (right axis) before (B) and after (A) 5000 AST cycles for a) in 1.0 KOH and for b) in 1.0 M EtOH and 1.0 M KOH in respect to initial metal loading. The experiments were performed with RDE configuration and constant Ar purge.

First, we discuss the dissolution behaviour of the catalysts in the EtOH-free electrolyte. Comparing Pd dissolution, it is clear that the highest dissolution is observed for Pd/C$_{AA}$, which is in line with the on-line ICP-MS study, and it is due to the larger exposed surface area. During the contact of the electrode with the electrolyte, almost 1 wt.% of the catalyst is dissolved. The 5000 cycles AST resulted in an additional 3-4 wt.% dissolution. This observation implies that for tiny Pd particles, Pd dissolution is possible at mild potentials ≤ 0.5 V$_{RHE}$. As for the synthesized catalysts, all of them revealed relatively high stability with dissolution below 1 wt.%. The most stable is the catalyst with Fe oxide. The somewhat higher Pd dissolution from PdSn/C and PdNb/C is likely related to the leaching of the secondary element. In the case of PdSn/C, since there is some alloying, the dissolution of Sn atoms can also carry over some dissolution of Pd. For Nb, no alloying was deduced from the physical characterization, and the lower stability could be due to the mixing of the two different oxide phases, which could induce dissolution in the contact areas between them. Indeed, while no Fe dissolution was observed, both Sn and Nb demonstrated low stability in this test. Interestingly, while Sn dissolved predominantly during the contact, the leaching of Nb was attributed to the AST cycles. Independent of the dissolution mode, 38 wt.% and 72 wt.% losses of Sn and Nb, respectively, render both elements impractical for real applications. Interestingly, despite the observed massive Sn and Nb dissolution, the EOR activity of the catalysts was still high and even increased. This can be explained by the dissolution of inactive species that do not contribute to the reaction.

To understand the dissolution behavior of all the studied elements, we focus on the thermodynamic data summarised by Pourbaix [51]. The Pourbaix diagrams can be found in the SI. In the case of Pd, no soluble species are shown to be stable in the Porbaix diagram at pH = 14, and only solid-solid transitions to form PdO and PdO$_2$ are found at ca. 0.9 V$_{RHE}$ and 1.25 V$_{RHE}$ for the alkaline pH values studied here. Therefore, the observed Pd dissolution must be mainly transient from forming these oxide species (and other sub-stoichiometric surface oxides formed at lower anodic potentials) and their subsequent reduction in the backward potential scans. The observed changes in stability between different samples are probably due to the influence of the second metal in these oxide formation and reduction processes and Pd particle size.

Regarding Sn, the transition between metallic Sn and subsequent soluble species ($HSnO_2^-$ and/or $SnO_3^{2-}$) takes place at ca. -0.25 $V_{RHE}$. According to thermodynamics, this means that the soluble species would be more stable in the studied potential ranges than the solid metallic or oxidic Sn. The fact that not all the Sn is dissolved indicates that the dissolution kinetics of the Sn atoms in the Pd-Sn alloy and the initially present $SnO_2$ is slow.

According to the Pourbaix diagram, Nb tends to form oxides ($Nb_2O_5$) in the investigated potential range. Consequently, its dissolution behaviour depends on the intrinsic properties of the particular layer oxide and its stability with respect to the working solution. Since no particular complexing agents are present in the 0.05 M KOH solution, one would not expect appreciable Nb dissolution according to the thermodynamics. The fact that significant dissolution is observed during the RDE experiments implies problems in the compactness and continuity of the initial compact layer, which are not considered in the classic thermodynamic treatment and would favor material dissolution. Moreover, as for Pd, transient Nb dissolution processes cannot be excluded.

Finally, concerning Fe, the stable species in the studied potential ranges in alkaline solutions are $Fe_3O_4$ and $Fe_2O_3$. In this case, no Fe dissolution was detected during the experiments. Therefore, the Fe oxide species in the prepared material possess adequate properties, leading to high stability against dissolution in agreement with the classical thermodynamic data.

The latter analysis, according to the Porurbaix diagrams, showcases that the stability of the different metals is determined by several factors that can have various importance depending on the particular element, such as transient dissolution from surface oxide formation and reduction processes, different stability of the soluble species and/or surface oxides in the studied potential range, or the specific quality of the formed oxide layer which cannot be addressed in the general thermodynamic treatments. In the case of $PdFe_3O_4/C$, less Pd dissolution was detected compared to the other investigated materials despite the very pronounced area observed in the cyclic voltammograms of the RDE experiments in the absence of fuel. Thus, the presence of Fe and its stability may contribute to the lower dissolution of Pd.[31,32].

Samples of the electrolytes obtained in the EOR experiments using RDE were taken before and after the AST studies to assess the dissolution of the metals present in the electrocatalysts in the presence of ethanol (**Fig. 5b**). In general, Pd dissolution is slightly higher in the presence of ethanol for all the electrocatalysts. Studies involving different fuels such as methanol, isopropanol, formic acid, and glycerol have not yet revealed a general/expected dissolution behavior. In some cases, the presence of the fuel does not affect the dissolution of the base metal but may contribute to the dissolution of the secondary metal [12,49,52,53]. In this study, the slightly most pronounced dissolution of Pd in the presence of ethanol can be attributed to the adsorption of alcohol molecules on the surface of the metallic Pd and suppression of Pd oxide formation required for passivation [12,43,53]. Still, it can be observed that this increase is minimal for PdSn/C, and it is also very small for PdFe$_3$O$_4$/C. Adding EtOH to the electrolyte does not change the relative dissolution behavior of Pd and secondary metals. Both Sn and Nb dissolve, while Fe is stable. However, the dissolution of Pd is slightly higher. Together with the previous observations, it can be concluded that PdSn/C and PdFe$_3$O$_4$/C would be the best electrocatalysts in terms of activity and stability among the ones studied here.

## 4. CONCLUSIONS

The dissolution stability of different Pd-based ethanol oxidation reaction electrocatalysts was investigated using in-situ and ex-situ ICP-MS approaches. Applying on-line SFC-ICP-MS, a snapshot of Pd, Sn, Nb, and Fe dissolution was obtained in a broad potential window. Moreover, the onset potentials for dissolution of Pd and second base elements were identified and related to the different physicochemical properties of the studied materials. It was found that, at least for the applied concentrations, ethanol does not affect the onset of Pd dissolution. AST studies were conducted in RDE configuration to evaluate electrocatalyst stability under simulated operating conditions, i.e., with cycles up to 0.5 V$_{RHE}$ in the absence and presence of ethanol. Despite the higher activity for PdFe$_3$O$_4$, this electrocatalyst presented less Pd dissolution when compared to the other studied materials, probably due to a reduced Pd transient dissolution kinetics as a result of the presence of the secondary metal and the fact that no detectable Fe dissolution, which could take out also Pd atoms, occurs. The PdSn/C material shows similar stability to Pd/C but has higher activity, making it also a suitable electrocatalyst towards the EOR. Higher Pd dissolution was observed in the presence of ethanol. Still, this difference is very small for PdSn/C and PdFe$_3$O$_4$/C, which also support the suitability of these materials to be used in ADLFCs. To sum up, the results showed in this work suggest that Sn- and Fe-based Pd nanoparticle electrocatalysts are promising candidates for ADLFC technology due to their improved activity and good stability performances. It is also highlighted that not only activity but also stability plays a key role when choosing proper electrocatalysts for ethanol oxidation in a practical device, and the parameters for the protocols employed for investigating this property, like, for example, the studied potential range, must be carefully selected.

**Acknowledgments**

The authors acknowledge financial support from the following Brazilian research financing institutions: Fundação de Amparo à Pesquisa do Estado de São Paulo (FAPESP, 2018/18675-8, 2021/10033-0, 2017/10118-0, 2017/21846-6, 2017/22976-0, 2017/26288-1, 2020/14100-0, 2022/15252-4) and Conselho Nacional de Desenvolvimento Científico e Tecnológico (CNPq, 308663/2023-3, 402609/2023-9).